\documentstyle[twocolumn,aps,prl,floats,epsf]{revtex}
\begin{document} 
\twocolumn[\hsize\textwidth\columnwidth\hsize\csname
@twocolumnfalse\endcsname
\draft
\title{Breathing Spots in a Reaction-Diffusion System} 
\vskip .1truein
\author{D. Haim$^{(1)}$, G. Li$^{(1)}$, Q. Ouyang$^{(1)}$,
W.D. McCormick$^{(1)}$, Harry L. Swinney$^{(1)}$, A. Hagberg$^{(2)}$
and E. Meron$^{(3)}$}

\vskip .2truein
\address{$^{(1)}$ Center for Nonlinear Dynamics and
Department of Physics, The University of Texas at Austin,\\ 
Austin, TX 78712\\ 
$^{(2)}$ Center for Nonlinear Studies and T-7, Theoretical Division,
Los Alamos National Laboratory, Los Alamos, NM 87545\\
$^{(3)}$ The Jacob Blaustein Institute for Desert Research and
The Physics Department, Ben-Gurion University\\
Sede Boker Campus 84990, Israel}

\date{15 May 1996}

\maketitle

\begin{abstract}
A quasi-2-dimensional stationary spot in a disk-shaped chemical reactor is 
observed to bifurcate to an oscillating spot when a control parameter is 
increased beyond a critical value. Further increase of the control 
parameter leads to the collapse and disappearance of the spot.  
Analysis of a bistable activator-inhibitor model indicates that 
the observed behavior is a consequence of interaction of the 
front with the boundary near a parity breaking front bifurcation. 
\end{abstract}


\vspace{-0.3in}
\vskip2pc]

\narrowtext


Oscillations in spatially extended chemical systems are often the
result of underlying oscillating dynamics of the local chemical
kinetics \cite{FiB:85}.  In systems with nonuniform spatial structures,
oscillations may also be driven by {\it diffusion}.  Spiral waves and
breathing spots in excitable and bistable media are examples of
oscillatory behaviors where the local kinetics without diffusion
converge to stationary uniform states, while spatial structures undergo
oscillations. Chemical spirals have been observed and studied for more
than two decades \cite{KaSh:95},
but breathing spots have not been previously observed although they have
been found in numerical and analytical studies of activator-inhibitor
models \cite{breathing,HaMe:94}.

Figure 1 shows an example of the breathing spots observed in our
study of a ferrocyanide-iodate-sulfite (FIS) reaction \cite{GaSh:87}
in a quasi-2-dimensional reactor. The breathing motion 
arises as a control parameter
is increased and an initially stable circular front (the spot boundary)
becomes unstable.  Further increase in control parameter eventually 
leads to the front rebounding from the cell boundary and propagating
inward until the spot collapses and disappears.
The breathing motion is interpreted as transitions between
left and right propagating fronts near a parity breaking front
bifurcation \cite{EHM:95}.  The rebound phenomenon leading to spot
collapse is attributed to 
crossing the front bifurcation as the control parameter is increased.  
We will first describe the experimental system and then present the
observations and the interpretation of the results in terms of a model
reaction-diffusion system.

\begin{figure}[htb]
\epsfxsize=3.5 truein  \centerline{\epsffile{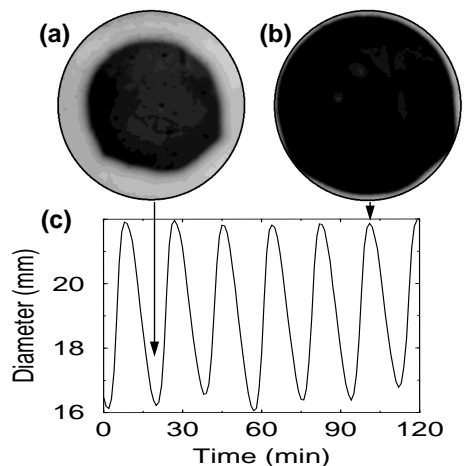}}
\caption{An oscillating circular spot at (a) minimal size and (b) maximal
size. (c) Time evolution of the spot diameter. The input concentrations to
the reservoir in contact with the gel reactor are
${\rm [H_2SO_4]} = 3.35 \;{\rm m}M$, ${\rm [IO_3^-]} = 75.0\;{\rm m}M$,
${\rm [SO_3^{-2}]} = 89\;{\rm m}M$,  and
${\rm [K_4Fe(CN)_6\cdot 3H_2O]} = 20 \;{\rm m}M$. The flow rate is 179 
ml/h.}
\label{fig1}
\end{figure}
 

The chemical patterns form in a thin gel layer that allows reaction and
diffusion processes but prevents convection.  The apparatus is similar
to that used by Lee {\it et al.} \cite{LeSw:95}. A polyacrylamide gel
layer (25 mm diameter, 0.3 mm thick) is in contact with a well-stirred
reservoir (2.8 ml volume) that is continuously fed with reagents of the
FIS reaction.  Reagents are fed first to
a premixer (1.0 ml volume) in two streams, one with $\rm{H_2SO_4}$ and
$\rm{NaIO_3}$ and the other with $\rm{Na_2SO_3}$ and
$\rm{K_4Fe(CN)_6\cdot 3H_2O}$.  The output of the premixer is fed to
a stirred reservoir that is in contact with the gel layer. The
reservoir diameter is 22 mm; thus the outer 1.5 mm width edge of the
gel is not in contact with the reservoir. The entire system is immersed
in a water bath maintained at $\rm{T=30^{\circ} C}$.  The side of the
gel opposite to the chemical reservoir is a window through which the
gel is illuminated with blue light (400-440 nm), and the patterns are
viewed using a video camera.  The system is studied in a range in
which the homogeneous FIS reaction has two stable states, one with low pH 
(about 4) and the other with high pH (about 7) \cite{GaSh:87,LeSw:95}. In the
observed patterns the black and white regions correspond to the low and
high pH states, respectively \cite {LeSw:95}.

We will now describe experiments that yield oscillating spots.  At low
flow rates the gel reactor is uniformly black; at high flow rates,
uniformly white \cite{LeSw:95}.  When the system is switched on at
intermediate flow rates, a black spot emerges in the center of the
reactor.  The spot is initially irregular but evolves to an circular
spot centered in the reactor. Above a critical flow rate (about 150
ml/h for the reagent concentrations used; see Fig.~1), the circular
spot oscillates in size as it approaches its asymptotic stationary
state.  Beyond a higher critical value of the flow rate (158 ml/h),
the circular spot becomes unstable and begins to oscillate in size, as
Fig.~2 illustrates. The oscillations are nearly sinusoidal just beyond
the onset of instability (Fig.~2(a)), while well beyond onset they
become relaxational (Fig.~2(b)). Measurements of the amplitude $A$ of
the oscillation as a function of flow rate indicate that the
transition is a Hopf bifurcation: $A^2$ increases linearly with
distance above transition, as shown in Fig.~2(c).

\begin{figure}[htb]
\epsfxsize=3.0 truein  \vskip .3in \centerline{\epsffile{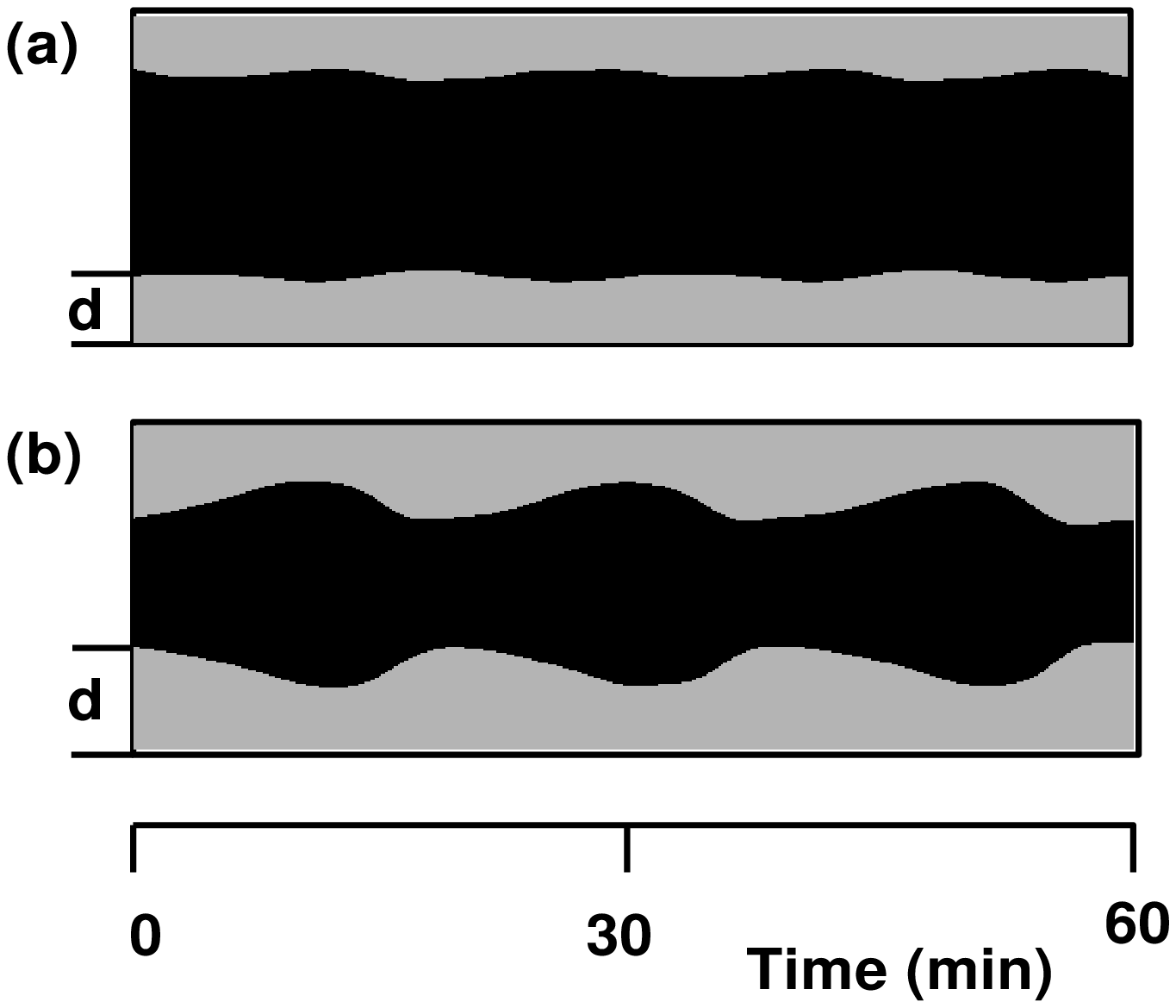}}
\epsfxsize=3.0 truein  \vskip .3in \centerline{\epsffile{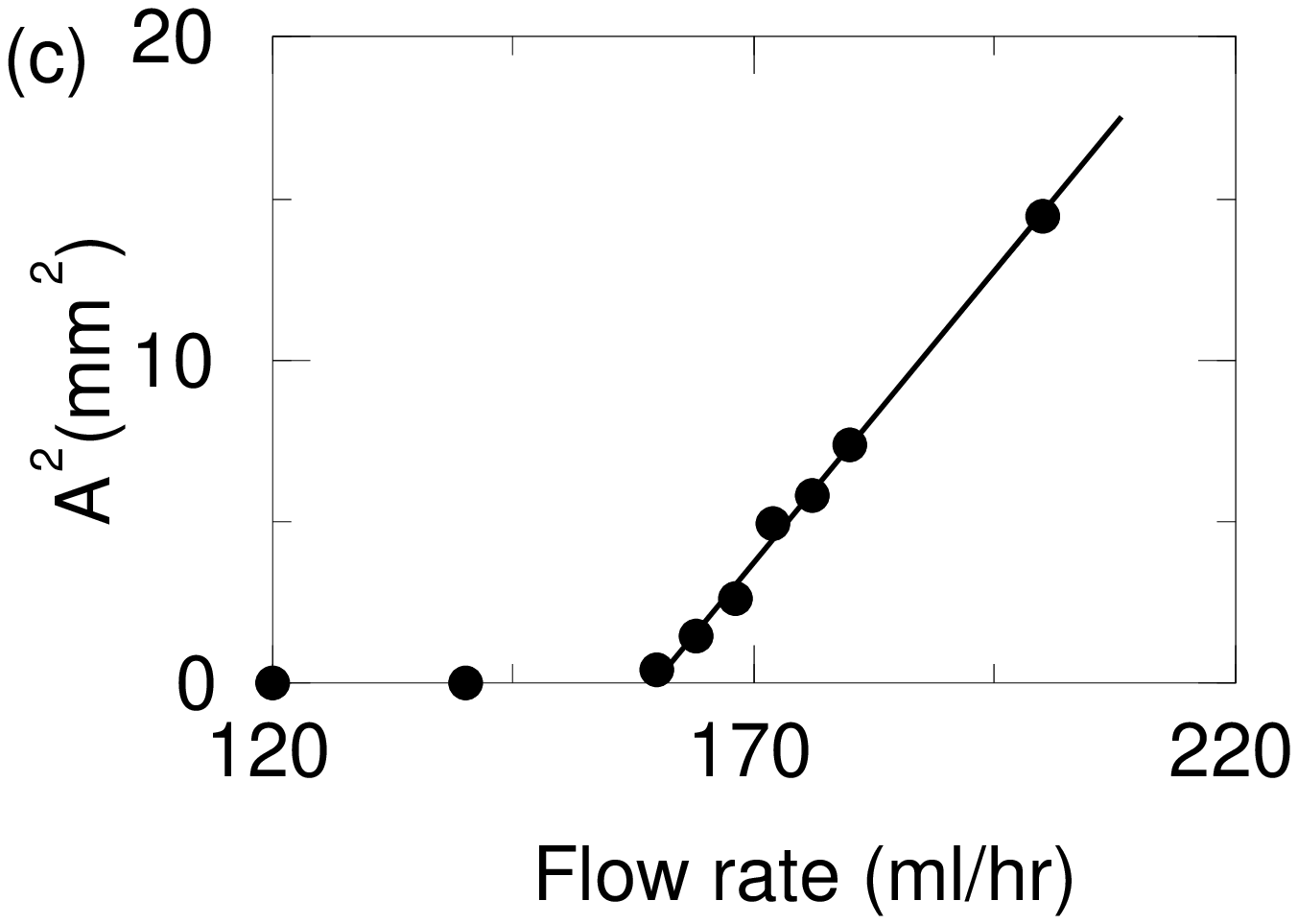}}
\caption{Time evolution of the cross section of a spot.
(a) Sinusoidal oscillations just beyond the onset of instability 
($F=160$ ml/h).
(b) Relaxation oscillations far beyond the instability onset ($F=200$ ml/h).
(c) The square of the spot oscillation amplitude as a function of
flow rate.}
\label{fig2} 
\end{figure}

Beyond a yet larger flow rate (260 ml/h), a shrinking spot does not
stop shrinking at a minimum size but instead continues to shrink until
the spot disappears. As the spot collapses, a new black ring emerges near
the outer edge of the reactor, creating two new fronts,
as Fig.~3(d) illustrates.  The inner front travels inward while the outer
front initially travels towards the boundary but then rebounds and
travels inward. The ring then collapses to a black spot that shrinks
and disappears. 
The space-time diagram in Fig.~3(d) shows that 
this whole process is periodic. 
\begin{figure}[htb]
\epsfxsize 4.5 truein  \vskip .3in \centerline{\epsffile{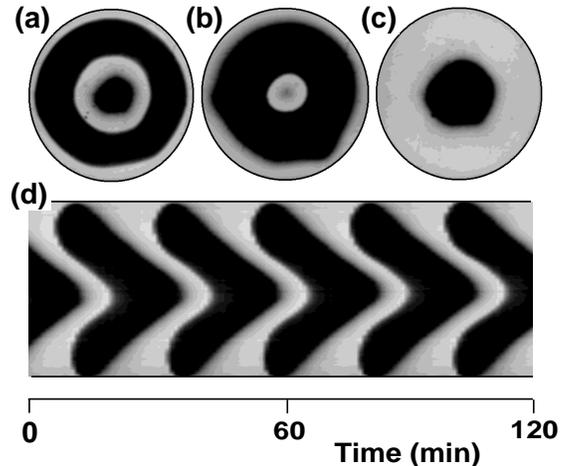}}
\caption{Periodic emergence of black rings near
the boundary leading to spot collapse near the center of the reactor.
(a)-(c) The dynamics of a single period, viewed at $80 min$, $85 min$,
and $93 min$, respectively; the time origin is arbitrary.
(d) The space-time evolution of a cross section of 
the reactor image. The flow rate is 280 ml/h.}
\label{fig3}
\end{figure}


We interpret these observations as interactions of the chemical front
with the reactor boundary in the vicinity of a parity breaking
front bifurcation.  The parity broken front states correspond to a
black state invading a white state (black-white front) and a white state
invading a black state (a white-black front).  Although the
two fronts connect the same states, they differ in their inner structures and 
consequently in their direction of propagation.
To develop this
interpretation we consider a model of a bistable reaction-diffusion 
system that exhibits pattern phenomenology similar to that observed in 
the FIS reaction \cite{HaMe:94b}, and is simpler for analysis than the
G\'asp\'ar-Showalter model for this reaction \cite{GaSh:87}.
The model equations in one space dimension are
\begin{mathletters}
\label{eq:model}
\begin{eqnarray}
u_t&=&u-u^3-v+u_{xx}\,,\label{eq:modelu} \\
v_t&=&\epsilon\,(u-a_1v-a_0)+\delta v_{xx}\,,
\end{eqnarray}
\end{mathletters}where the subscripts $x$ and $t$ denote partial
derivatives. For a fixed $a_1>1$ there is a parameter range
where the system is
bistable; it has two coexisting stable stationary uniform
states $(u_+,v_+)$ and $(u_-,v_-)$. The parameter 
$a_0$ will be associated with the flow rate $F$, and the two states
$(u_+,v_+)$ and $(u_-,v_-)$ with the white (high pH) and black
(low pH) states, respectively. 

For $a_0=0$ and fixed $\delta$ the system (\ref{eq:model})
exhibits a pitchfork front bifurcation (also known as the nonequilibrium 
Ising-Bloch (NIB) bifurcation) as the parameter $\epsilon$
is decreased past a critical value, $\epsilon_c$. For
$\epsilon>\epsilon_c$ there is a single, stationary front solution 
connecting $(u_+,v_+)$ at $x=-\infty$ to $(u_-,v_-)$ at
$x=\infty$.  At $\epsilon=\epsilon_c$ the stationary front solution
becomes unstable and a pair of counterpropagating front solutions 
with velocities $c\propto \pm \sqrt{\epsilon_c-\epsilon}$ 
appear. These are the parity broken front states corresponding to the 
black-white ($c<0$) and the white-black ($c>0$) fronts.
When $a_0\ne 0$ the pitchfork bifurcation becomes imperfect, i.e.,
unfolds into a saddle node bifurcation where, at
$\epsilon=\epsilon_c(a_0)$, a stable-unstable pair of front
solutions appears in addition to the stable front solution that
already exists.

The front bifurcation can also be traversed by varying other
parameters, in particular by increasing $a_0<0$.  
We will investigate the effect of a no-flux boundary on the
dynamics of a front as $a_0$ is increased.  
Using a singular perturbation approach with $\epsilon/\delta$ as a small
parameter, we derive a relation between the front velocity, $c$,
and the distance from the front to the boundary, $d$.
The Hopf bifurcation observed in the experiment will be
associated with the $c-d$ relation becoming multivalued.

In a moving coordinate system (1) becomes 
\begin{mathletters}
\label{eq:inner}
\begin{eqnarray}
u_{zz}+c u_z+ u-u^3-v&=&0\,,\\
\delta v_{zz}+c v_z+\epsilon\,(u-a_1v-a_0)&=&0\,, 
\end{eqnarray}
\end{mathletters} where $x_f(t)$ is the position of the narrow front 
structure, $z=x-x_f(t)$, and $c=\dot x_f$ is the front velocity. 
The boundary conditions are
$(u,v)\to(u_-,v_-)$ as $z\to\infty$ and $(u_z,v_z)=(0,0)$ at $z=-d$, where
$d$ is the distance from the front to the boundary. In obtaining 
(\ref{eq:inner}) we 
assume that the front velocity $\dot x_f=c$ is small so that the explicit 
time dependence in the moving frame can be neglected. 
The front velocity can be controlled by varying $a_1$; increasing $a_1$
leads to lower velocities. 

We first solve equations (\ref{eq:inner}) in the front, or
``inner'', region. Letting $\mu=\epsilon/\delta\to 0$ at finite
$\eta=\sqrt{\epsilon\delta}$, we obtain the equation
$u_{zz}+cu_z+u-u^3-v_f=0$, subject to the boundary conditions
$u\to u_+(v_f)$ as $z\to-\infty$ and $u\to u_-(v_f)$ as $z\to\infty$.
Here, $v_f$ is the value of $v$ at the front, and
$u_\pm(v_f)$ are the largest and smallest roots of $u-u^3-v_f=0$.
Solving the inner problem yields
\begin{equation}
v_f=-\frac{\sqrt{2}}{3}c\,. \label{eq:innervf} 
\end{equation}

Another relation between $c$ and $v_f$ is obtained by solving the
equations in the regions to the left and to the right of the
front, the ``outer'' regions. Rescaling the coordinate system
according to $\zeta=\sqrt\mu z$ and letting $\mu\to 0$ gives
\cite{HaMe:94}
\begin{eqnarray}
v_{\zeta\zeta}+\frac{c}{\eta}v_\zeta-q^2(v-v_+)&=&0\qquad\zeta<0\,,\nonumber\\
v_{\zeta\zeta}+\frac{c}{\eta}v_\zeta-q^2(v-v_-)&=&0\qquad\zeta>0\,, \nonumber
\end{eqnarray}
with the boundary conditions, $v(0)=v_f$, $v(\infty)=v_-$,
and $v_\zeta(-\sqrt{\mu} d)=0$. Here, $q^2=a_1+1/2$ and 
$v_\pm=(\pm 1-a_0)/q^2$.
Solving this boundary value problem and matching the derivatives of $v$ at
$\zeta=0$ we find
\begin{equation}
v_f=-{1\over q^2} 
\left[ 
\frac{c}{\alpha} +a_0 + 
\left( 1-\frac{c}{\alpha} \right)
e^{-\alpha d/\delta}
\right] 
\,,\label{eq:outervf}
\end{equation}	
where $\alpha=\sqrt{c^2+4\epsilon\delta q^2}$. 

Equating (\ref{eq:innervf}) and (\ref{eq:outervf}) gives an
implicit relation between $c$ and $d$.  
In the limit $d\to\infty$ this relation reproduces
the front bifurcation line \cite{HaMe:94}, which we may write as
$a_0=a_{0b}(\epsilon)$.  Fig.~4(a) shows a graph of the $c-d$ relation 
far into the single front regime; Fig.~4(b), near the front 
bifurcation; and Fig.~4(c), beyond the bifurcation.
The figures also show trajectories
representing the front dynamics as obtained by direct numerical
solution of equations (\ref{eq:model}).
\begin{figure}[htb]
\vspace{4.5in}
\includegraphics{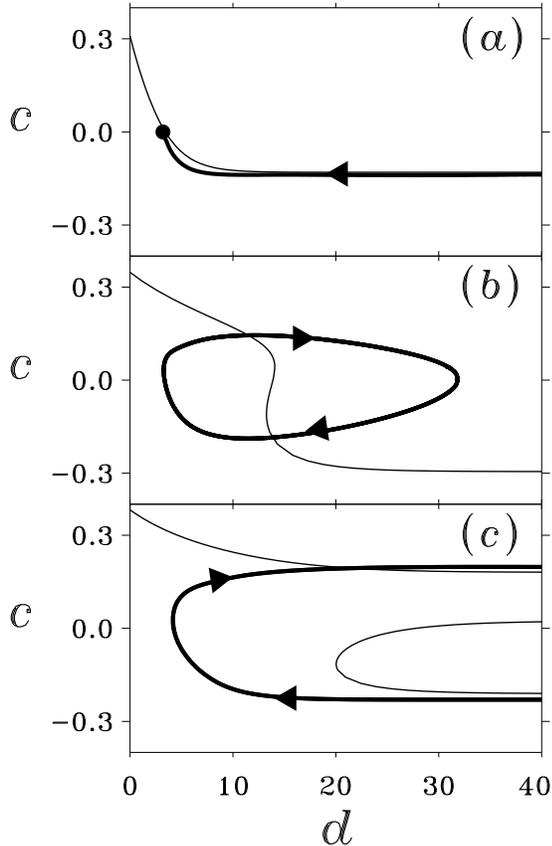}
\caption{The dynamics of fronts near a boundary.  The thick lines
represent the front distance from the boundary, $d$,
and front speed, $c$, computed from
the numerical solution of equations  (\protect\ref{eq:model}). 
The thin lines are the solutions to equations (\protect\ref{eq:innervf}) 
and (\protect\ref{eq:outervf}). 
(a) Far into the single front regime, a front approaching from
large $x$ values stops at a fixed distance from the boundary;
$a_0=-0.2$, $\epsilon=0.025$.
(b) Near the front bifurcation the
$c-d$ relation is multivalued and the front begins oscillating;
$a_0=-0.1$, $\epsilon=0.0025$.
(c) After crossing the front bifurcation the upper branch of
the $c-d$ relation exists for all $d$ and the approaching front
rebounds and propagates away from the boundary to infinity;
$a_0=-0.01$, $\epsilon=0.0025$.
In all three cases, $a_1=5$ and $\delta=2.0$.
}
\label{fig4}
\end{figure}


The monotonic velocity relation and the corresponding trajectory in
Fig.~$4(a)$ describe the approach of a black-white front from the far
right (large $x$) to the boundary at $x=x_f-d$.  The solution converges
to a stationary front at some $d=d_0$. The negative slope at
$(c,d)=(0,d_0)$ implies stability of the stationary solution.
We associate with this scenario the formation of a black spot
observed at low flow rates in the experiment.

As $a_0$ is increased, the slope of the $c-d$ relation at
$(c,d)=(0,d_0)$ increases in absolute value and at a critical point,
$a_{0c}=-1+2q^3\sqrt{2\epsilon\delta}/3 $, diverges to infinity. Beyond
$a_{0c}$ the slope is positive and the $c-d$ relation is multivalued in
some range of distances from the boundary (Fig.~$4(b)$). The critical
point $a_{0c}$ corresponds to the onset of oscillatory front motion,
and explains  the Hopf bifurcation to a breathing spot observed in the 
experiment.

The oscillations can be regarded as periodic transitions between
left and right propagating fronts represented by the upper and lower branches
of the $c-d$ relation in Fig.~4$(b)$ \cite{HBKR:95}.  
The dynamics actually do not follow these branches
because we have neglected the explicit time dependence of the $u$ and
$v$ fields in the moving frame.  Near the bifurcation, $v_f$ becomes an
active degree of freedom responsible for transitions between the fronts
\cite{HMRZ:95}. The present analysis, however, accurately predicts
(within 3\% for $a_1=5$) the onset of breathing motion and
describes the dynamics far from the front bifurcation 
(see Fig.~$4(a)$ and $4(c)$).

Beyond the front bifurcation, $a_0>a_{0b}(\epsilon)$, the
upper branch in the $c-d$ relation extends to infinity, i.e.,
exists for all distances $d$. As Fig.~$4(c)$ demonstrates,
this type of $c-d$ relation results in the rebound of a front
approaching the boundary along the lower branch (black-white
front) and the escape to infinity along the upper branch
(white-black front). This behavior together with the circular
geometry of the experimental apparatus explain the rebound
and collapse of black spots observed at high flow
rates. 
Possible instabilities to transverse perturbations 
\cite{LeSw:95,HaMe:94b} are
damped in the present experiment by high curvature (small spots) and 
by the interaction with the circular reactor boundary (large spots).
The observation of traveling rings in the same parameter range as the
collapsing spots (Fig.~3) provides further evidence for this
interpretation since crossing the front bifurcation is
associated with the appearance of traveling waves
\cite{HaMe:94}. The periodic production of waves near the boundary might
be due to heterogeneities, which become significant near the oscillatory
regime at high flow rates, or global coupling effects. The latter
possibility is not highly probable as discussed below.

We have been able to explain breathing spots in terms of dynamic front
transitions near a parity breaking front bifurcation.  The generic
nature of this bifurcation suggests that similar behavior can be
expected in other systems with fronts.  Breathing spots can also
arise from global coupling, as found by Middya and Luss in model
equations \cite{MiLu:95}.  Indeed, we find that decreasing the
strength of a global coupling term added to (1a) leads to a scenario
similar to that observed: stationary, oscillating, and collapsing
spots.  In our experiment global coupling can arise from diffusion of
chemicals from the gel back into the stirred reservoir.  To test
whether such coupling is significant, we monitor the pH in the
reservoir.  We find that, despite the oscillations in the gel, the pH
in the reservoir is time-independent.  This suggests
that the primary mechanism
leading to the oscillating spots is not global coupling but the
one presented in this Letter:
interactions of the front with the boundary near a parity 
breaking bifurcation.

This work was supported by the U.S. Department of Energy Office of Basic
Energy Sciences and the Robert A. Welch Foundation.
E.~M. acknowledges the support of the Israel Ministry of Science and the Arts.


\end{document}